\documentclass[aps,prl,twocolumn,amsmath,superscriptaddress,reprint]{revtex4-1}

\usepackage{graphics}
\usepackage{amsfonts}
\usepackage{bbm}

\begin{document}

%\preprint{}

\title{Backscattering Differential Ghost Imaging in Turbid Media}


\author{M. Bina}
\email[]{matteo.bina@gmail.com}
\affiliation{Dipartimento di Scienza e Alta Tecnologia, Universit\`{a} degli Studi dell'Insubria, Via Valleggio 11, I-22100 Como, Italy.}
\author{D. Magatti}
\altaffiliation{Present address: Light in Light s.r.l. via Ferrari 14, I-22100 Como, Italy.}
\affiliation{Dipartimento di Scienza e Alta Tecnologia, Universit\`{a} degli Studi dell'Insubria, Via Valleggio 11, I-22100 Como, Italy.}
\author{M. Molteni}
\affiliation{Dipartimento di Scienza e Alta Tecnologia, Universit\`{a} degli Studi dell'Insubria, Via Valleggio 11, I-22100 Como, Italy.}
\author{A. Gatti}
\affiliation{Istituto di Fotonica e Nanotecnologie - CNR, P.zza Leonardo da Vinci 32, Milano, Italy}
\affiliation{Dipartimento di Scienza e Alta Tecnologia, Universit\`{a} degli Studi dell'Insubria, Via Valleggio 11, I-22100 Como, Italy.}
\author{L. A. Lugiato}
\affiliation{Dipartimento di Scienza e Alta Tecnologia, Universit\`{a} degli Studi dell'Insubria, Via Valleggio 11, I-22100 Como, Italy.}
\author{F. Ferri}
%\email[]{fabio.ferri@uninsubria.it}
\affiliation{Dipartimento di Scienza e Alta Tecnologia, Universit\`{a} degli Studi dell'Insubria, Via Valleggio 11, I-22100 Como, Italy.}


\date{\today}

\begin{abstract}
In this Letter we present experimental results concerning the retrieval of images of absorbing objects immersed in turbid media via differential ghost imaging (DGI) in a backscattering configuration. The method has been applied, for the first time to our knowledge, to the imaging of small thin black objects located at different depths inside a turbid solution of polystyrene nanospheres and its performances assessed via comparison with standard  imaging techniques. A simple theoretical model capable of describing the basic optics of DGI in turbid media is proposed. 
\end{abstract}

%\keywords{}

\maketitle

%%%%%%%%%%%%%%%
%%% INTRODUCTION %%%
%%%%%%%%%%%%%%%

Ghost imaging (GI) is an optical technique for the retrieval of images via intensity correlation of two correlated light beams. The first experimental approach and theoretical explanation of GI was quantum-like \cite{Shih_first}, but after a long-standing debate \cite{Refs_debate}, it was finally demonstrated that GI can be also realized with classical light beams \cite{Ferri_PRL_2005,Shapiro_Gaussian}. Thermal GI, for instance, is performed with two spatially correlated speckle beams obtained by using a rotating ground glass and a beam splitter. The object beam illuminates the object and is collected by a bucket detector with no spatial resolution, while the reference beam is recorded by a spatial-resolving detector, for example by a charge coupled device (CCD) camera. Recent improvements of the GI protocol have been achieved via computational GI that uses computer controlled spatial light modulators \cite{Shapiro_Computational}, compressive sensing GI where the algorithm for the data analysis benefits from the sparsity properties of the object \cite{Katz-CGI} and via Differential Ghost Imaging (DGI), which has been shown to perform much better than conventional GI when imaging weakly absorbing objects \cite{DGI}.\\
\indent The potentialities of GI with respect to standard (not correlated) imaging resides in its ability of forming images without necessity of any pixelated detector placed nearby the object. Thus GI is a good candidate for imaging objects immersed in optically harsh or noisy environments such as, for example, in a turbid medium or in the presence of optical aberrations. Recent applications of GI in this direction include imaging in presence of atmospheric turbulence  \cite{AtmosphericGI}, fluorescent ghost imaging \cite{FGI} and transmission GI in scattering media \cite{Gong-Han}. All these works have raised the very interesting debate whether GI is intrinsically more powerful than standard imaging and can be used, for example, as a standoff sensing technique which is \textit{immune} from atmospheric turbulence \cite{Meyers-Shapiro}.\\
\indent Following this debate, we propose for the first time in this letter the use of DGI for the imaging of absorbing objects immersed in a turbid medium, in proximity of its surface. We adopt a backscattering configuration of the bucket light detection similar to the schemes used in biomedical tissue imaging and we are able to provide information on the transmittance of the object as a function of its depth inside the turbid medium. We also  compare the performances of DGI with standard techniques, where the imaging is performed with a lens and a CCD, showing that the two methods are fairly equivalent.\\
\indent The experimental setup 
%%%%%%%%%%%%%%%%%%%
%%% THEORETICAL MODEL %%%%
%%%%%%%%%%%%%%%%%%%
\begin{figure}[b]
\resizebox{0.9\columnwidth}{!}{\includegraphics{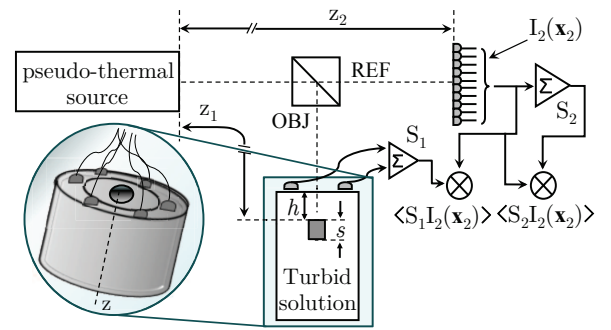} }
\caption{\label{Fig1} (Color online) Setup for backscattering DGI. The zoom shows a detail of the scattering cell with the six photodiodes used for the bucket detection.}
\end{figure}
for the DGI configuration is sketched in Fig.~\ref{Fig1}. The pseudothermal source, operating at $\lambda=0.532\mu$m, produces a rectangular collimated beam of \emph{deep Fresnel} speckles \cite{DeepFresnel} with a constant transversal size $\delta_x\simeq 82 \mu$m and longitudinal coherence length $\delta_z\sim \pi\delta_x^2/\lambda\simeq 40$mm. The beam area is $A_{b}=46.6\text{mm}^2$ and contains $N_{\text{speckle}}\simeq 6900$ speckles of coherence area $A_{\text{coh}}=\delta_x^2$ \cite{FerriLongitudinal}. The reference beam intensity $I_2({\bf{x}}_2)$ is recorded at a distance $z_2\simeq250$mm from the source by a CCD camera with pixel size $6.67\mu\text{m}\ll \delta_x$.   
The intensity $I_1({\bf{x}}_1)$ hits the object at a distance $z_1=z_2$ and is collected with a bucket detection in backscattering. The object, characterized by a spatial transmittance $T({\bf{x}})$ over the same area of the beam, is immersed in a turbid solution contained in a cylindrical cell (diameter $d=44$mm, length $L=60$mm) and it is allowed to move along the optical axis. The turbid solution is made of an aqueous solution of poly-disperse silica particles (Ludox PW-50, average particle diameter $\simeq 50$nm). Three volume fractions, $\phi_1=3.1\times 10^{-3}$, $\phi_2=7.8\times 10^{-3}$ and $\phi_3=15.6\times 10^{-3}$ were used, with corresponding transport mean free paths $l^*_1\simeq 17.3$mm, $l^*_2\simeq 6.7$mm and $l^*_3\simeq 3.8$mm. 
The light transmitted by the object and backscattered by the medium is collected by six photodiodes placed in a ring configuration (ring radius = 15mm) outside the cell around the object. Such a configuration ensures that the average output signal from the six photodiodes can be used as an effective bucket detector. Indeed, the transport mean free paths of our medium are much smaller than the average contour length $L_c$ that photons travel from the injection point to the escaping point at the photodiodes positions. Thus, thanks to the backscattering detection scheme, the light reaching the photodiodes is completely randomized and the measured signal is proportional to the overall power injected into the solution and transmitted by the object. This implies that, in a blank measurement with no object, $S_1^{\text{blk}}=\alpha S_2$, where  $S_2=\int_{A_b} I_2({\bf{x}_2}) d {\bf{x}_2}$ and $\alpha$ is a factor which takes into account any unbalancing (beam splitter, detectors, random medium) between the two arms. 
\begin{figure}[t]
\resizebox{0.95\columnwidth}{!}{\includegraphics{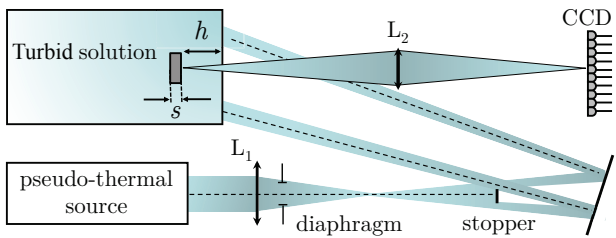} }
\caption{\label{ImSetUp} (Color online) Setup for standard imaging. The object is back illuminated with a ring of speckled light shaped by using the lens $L_1$, a diaphragm and a stopper. The lens $L_2$ realizes a 1:1 imaging of the object on the CCD sensor.}
\end{figure}

The ghost imaging data analysis is carried out by using the DGI algorithm \cite{DGI} based on the measurement of the observable
\begin{equation}\label{O-}
\langle O_-({\bf{x}_2})\rangle=\langle S_1I_2({\bf{x}_2})\rangle - \frac{\langle S_1\rangle}{\langle S_2\rangle}\langle S_2I_2({\bf{x}_2})\rangle
\end{equation}
where $S_1$ and $S_2$ are the bucket signals collected in the object and reference arms, respectively, and $\langle\,\cdots\rangle$ is performed over independent speckles configurations. From Eq.~(\ref{O-}) we measure the fluctuations of the transmission function $\delta T_m({\bf{x}})$, while the measured spatial average of the transmission function is $\overline{T}_m=\langle S_1 \rangle/(\alpha\langle S_2\rangle) $. The measured transmittance of the object is thus computed as $T_m({\bf{x}})=\overline{T}_m+\delta T_m({\bf{x}})$.\\ 
\indent The recovered ghost images are compared with standard imaging measurements performed with the setup shown in Fig.~\ref{ImSetUp}. This setup is in a sense the reverse of the GI one since illumination is performed backward and detection forward, and is similar to the ones used in biomedical tissue imaging. A ring of speckled light, formed by reshaping the speckle beam with a diaphragm and a stopper, diffusively illuminates the object from the back satisfying the turbid medium condition  $l^*\ll L_c$.  A macro objective (Nikon AF Micro Nikkor 60mm f/2.8D) realizes the imaging of the object onto the CCD sensor with a 1:1 magnification.\\
\indent In our experiments we considered two simple objects characterized by a binary transmission function, with $T({\bf{x}})=0,1$. 
\begin{figure}[t]
\resizebox{0.9\columnwidth}{!}{\includegraphics{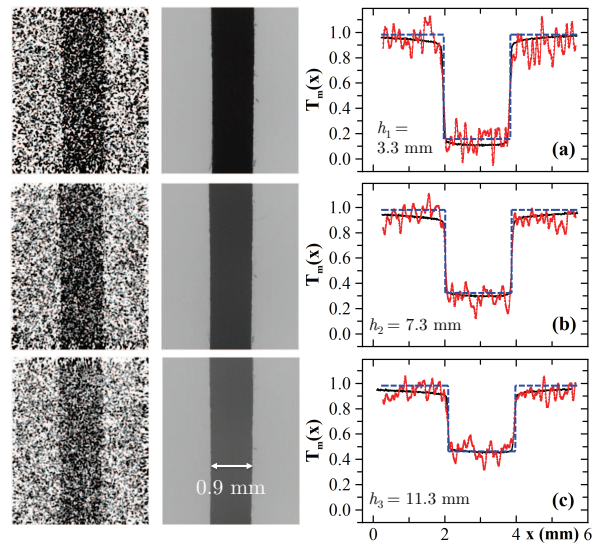} }
\caption{\label{Fig3} (Color online) Images of a thin black cardboard (section 1.8mm $\times$ 8mm, thickness $s\simeq 400\,\mu$m) inside a turbid solution with $l^*=17.3$mm at $h_1=3.3$mm (a), $h_2=7.3$mm (b) and $h_3=11.3$mm (c), obtained with DGI (left column) and standard imaging (central column). The right column plots the corresponding horizontal sections averaged over the vertical dimension of the image: DGI (red curve), standard imaging (black curve), theoretical model (blue dashed line).}
\end{figure}
The first object was a thin black cardboard of section 1.8mm $\times$ 8mm and thickness $s\simeq 400\,\mu$m. The turbid solution with  $l^*=17.3$ mm was used. Fig.~\ref{Fig3} reports three examples of images recovered via DGI (left column) and standard imaging (central column), together with their horizontal sections averaged over the vertical dimension of the image (right column). The DGI images were obtained by averaging 6000 independent speckle configurations. The figure shows that, as $h$ is increased, the visibility of recovered images (both standard and DGI) becomes smaller because the central part of the image, where the object is totally absorbing ($T=0$), becomes increasingly transmissive, passing from $T_m\simeq 0.1$ ($h=3.3$mm) to $T_m\simeq 0.5$ ($h=11.3$mm).
The figure shows also that the matching between DGI and standard imaging is excellent, although, as expected, DGI suffers of a much lower SNR. The latter one can be easily improved \cite{DGI} by increasing the number of measurements. The blue dashed lines in the right columns are the result of a simple model for DGI described below.\\ 
\begin{figure}[t]
\centering
\resizebox{0.91\columnwidth}{!}{\includegraphics{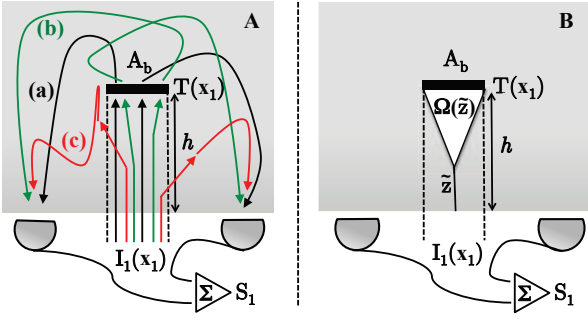} }
\caption{\label{Scheme_S1} (Color online) (A): backscattering scheme showing that the bucket signal $S_1$ is composed by three main contributions: (a) straight light hitting the object at a depth $h$, (b) forward scattered light that illuminates the object and (c) scattered light collected without intercepting the object. (B): scheme for the fraction $\varepsilon_h$ of the light scattered in the forward direction that hits the object.}
\end{figure}\indent Under the assumption 
%%%%%%%%%%%%%%%%%%%
%%% THEORETICAL MODEL %%%%
%%%%%%%%%%%%%%%%%%%
of a thin object of thickness $s$, located at a depth $h$ from the entrance face of the cell such that $s\ll h \ll l^*$, the light travelling along the distance $h$ can be considered as undergoing only single scattering events. Thus, the light that hits the object (as sketched in Fig.~\ref{Scheme_S1}(A)) is made of two main contributions: (\ref{I1h_a}) the straight non-scattered light that reaches the object with a probability $\beta_h=\exp(-h/l^*)$ given by the Lambert-Beer (L-B) law \cite{VanDeHulst} and (\ref{I1h_b}) the light that, after being scattered with probability $1-\beta_h$, reaches the object with probability $\varepsilon_h$ determined by geometrical factors. Hence, we may write this intensity as 
\begin{subequations}\label{I1h}\begin{align}
I_1^{(h)}({\bf{x}_1})&=\beta_h I_1({\bf{x}_1})\;+\label{I1h_a}\\
&+\big (1-\beta_h\big )\varepsilon_h\frac{\int_{A_b}I_1({\bf{x}_1'})d{\bf{x}_1'}}{A_b}i_{sc}({\bf{x}_1})\label{I1h_b}
\end{align}\end{subequations}
where $\int_{A_b}I_1({\bf{x}_1'})d{\bf{x}_1'}/A_b$ is the incident average intensity and $i_{sc}({\bf{x}_1})$ is the distribution of the scattered light, totally uncorrelated to $ I_1({\bf{x}_1})$, normalized so that $ \int_{A_b}\langle i_{sc}({\bf{x}_1})\rangle d{\bf{x}_1}/A_b=1$. The bucket signal $S_1$ is given by $\int_{A_b} I_1^{(h)}({\bf{x}_1}) T({\bf{x}_1})d{\bf{x}_1}$ plus a third contribution coming from the scattered light that does not pass through the object. $S_1$ can be written in the following way
\begin{subequations}\label{S1}\begin{align}
&S_1\propto \;\Big \{\beta_h\int_{A_b} T({\bf{x}_1}) I_1({\bf{x}_1}) d {\bf{x}_1}+ \label{S1_a}\\
&\big (1-\beta_h\big ) \varepsilon_h\frac{\int_{A_b}I_1({\bf{x}_1'})d{\bf{x}_1'}}{A_b}\int_{A_b} i_{sc}({\bf{x}_1}) T({\bf{x}_1})d{\bf{x}_1}+ \label{S1_b}\\
&\big (1-\beta_h\big ) \big (1-\varepsilon_h\big )\int_{A_b} I_1({\bf{x}_1}) d {\bf{x}_1} \Big \}\label{S1_c}.
\end{align}\end{subequations}
Note that if the object is placed on the surface ($h=0$), $\beta_0=1$, we recover the common definition of the bucket signal (\ref{S1_a}) used in absence of the turbid medium, regardless of $\varepsilon_h$. The dimensionless factor $\varepsilon_h$ reads
\begin{equation}\label{epsilon}
\varepsilon_{h}=\frac{1}{h}\int_0^h d\tilde{z}\, \beta_{\tilde{z}}\, \omega(\tilde{z},h)
\end{equation}
which is an average along the object depth $h$ of the probability $\omega(\tilde{z},h)$ that scattered light, at a position $\tilde{z}$ from the surface of the cell, hits the object, weighted by the L-B factor $\beta_{\tilde{z}}$ (see Fig.~\ref{Scheme_S1}(B)). This probability corresponds to the fraction of light scattered within a maximum solid angle $\Omega(\tilde{z})$ subtended by the object (with area $A_b$), normalized to $\frac{8\pi}{3}=\int_{4\pi}\sin^2\phi\, d\Omega$. We assume Rayleigh scattering with an incident polarized electric field that forms an angle $\phi$ with the scattering direction. The integral in Eq.~(\ref{epsilon}) is computed numerically.\\ 
\indent Combining Eq.~(\ref{O-}) and Eqs.~(\ref{S1}), with the assumption of uniform illumination ($\langle I_1({\bf{x}_1}) \rangle=\langle I_1\rangle$, $\langle I_2({\bf{x}_2}) \rangle=\langle I_2\rangle$), and taking into account that $I_2({\bf{x}_2})$ and $i_{sc}({\bf{x}_1})$ are uncorrelated $\big (\langle I_2({\bf{x}_2})i_{sc}({\bf{x}_1}) \rangle=\langle I_2({\bf{x}_2}) \rangle \langle i_{sc}({\bf{x}_1}) \rangle= \langle I_2 \rangle\big )$, we derive an expression for the measured transmittance of the object $T_m({\bf{x}})$ in terms of the real $T({\bf{x}})$, which reads
\begin{equation}\label{Tm}
T_m ({\bf{x}})=\beta_h T({\bf{x}})+(1-\beta_h)\Big [ \varepsilon_h\overline{T} +1-\varepsilon_h\Big ]
\end{equation}
where $\overline{T}=\int_{A_b}T({\bf{x}_1}) d {\bf{x}_1}/A_b$ is the spatially averaged transmittance of the object.
As expected, Eq.~(\ref{Tm}) predicts that,  in the case of non-turbid media or in the case of objects placed at the surface of  the cell, whenever $h/l^*\to 0$, $T_m({\bf{x}})\to T({\bf{x}})$. But remarkably, although based on the assumption that $h\ll l^*$, Eq.~(\ref{Tm}) predicts also the correct behavior of $T_m({\bf{x}})$ for highly turbid media or objects deeply inside the scattering cell ($h/l^*\to\infty $), for which $\beta_h\to 0$ and $\varepsilon_h\to 0$. In these cases, indeed, the object becomes invisible and, consistently, Eq.~(\ref{Tm}) predicts $T_m({\bf{x}})\to1$. When Eq.~(\ref{Tm}) is applied to analysis of the images of Fig.~\ref{Fig3} (blue dashed lines in the third columns), the agreement with the experimental data is excellent in correspondence of the absorbing zones ($T({\bf{x}})=0$) of the object, while is somewhat less accurate for the transmissive zones ($T({\bf{x}})=1$). Overall, the simple model of Eqs.~(\ref{I1h},\ref{S1}) is able to capture the essential physics of DGI in turbid media. 
\begin{figure}[t]
\resizebox{0.83\columnwidth}{!}{\includegraphics{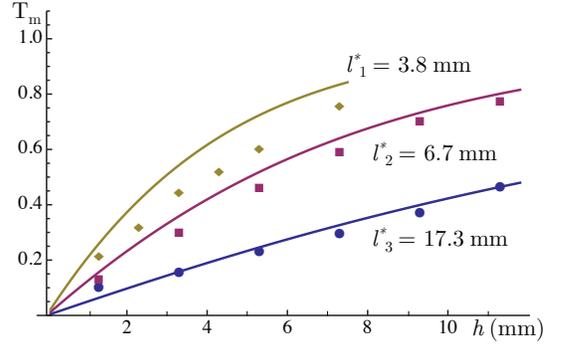} }
\caption{\label{Fig5} (Color online) Plots of the experimental data (symbols, whose dimensions are equivalent to error bars) and theory (solid curves) for the absorbing region of the object $T_m$ as a function of its depth $h$ for three different turbid solutions.}
\end{figure}
A more quantitative analysis of the data of Fig.~\ref{Fig3} is reported in Fig.~\ref{Fig5}, where we compare, as a function of $h$, the behaviors of the expected (Eq.~(\ref{Tm})) values $T_m$ of the absorbing zone ($T({\bf{x}})=0$, solid curves) with the experimental data, for the three solutions with $l^*_1\simeq 17.3$mm, $l^*_2\simeq 6.7$mm and $l^*_3\simeq 3.8$mm.
The agreement between theory and experiment is quite good for the $l^*=17.3$mm curve but becomes less accurate at higher turbidities, where the presence of increased multiple scattering reduces the validity of the assumptions used in the model of Eqs.~(\ref{I1h},\ref{S1}).\\
\indent Our results were also validated by measuring an absorbing sphere much smaller (diam = $0.9$mm) than the beam area ($\overline{T}\simeq 0.99$) in a turbid solution with $l^*=6.7$mm. Figure~\ref{Fig6} reports the images retrieved with DGI (left column) and standard imaging (central column), together with their corresponding radial profiles (right column) obtained averaging the images over the azimuthal angle. We notice that, as for Fig.~\ref{Fig3}, the object becomes less visible as $h$ is increased. In correspondence to the absorbing region of the object, we obtain values that ranges from $T_m\simeq 0.30$ ($h_1=2.4$mm) to $T_m\simeq 0.61 $ ($h_3=6.4$mm), a result that is equivalent for both DGI and standard imaging. The agreement between experimental results and the theoretical model is excellent, as shown in the third column of the figure.
%%%%%%%%%%%%%%%
%%% CONCLUSIONS %%%
%%%%%%%%%%%%%%%

\begin{figure}[t]
\centering
\resizebox{0.9\columnwidth}{!}{\includegraphics{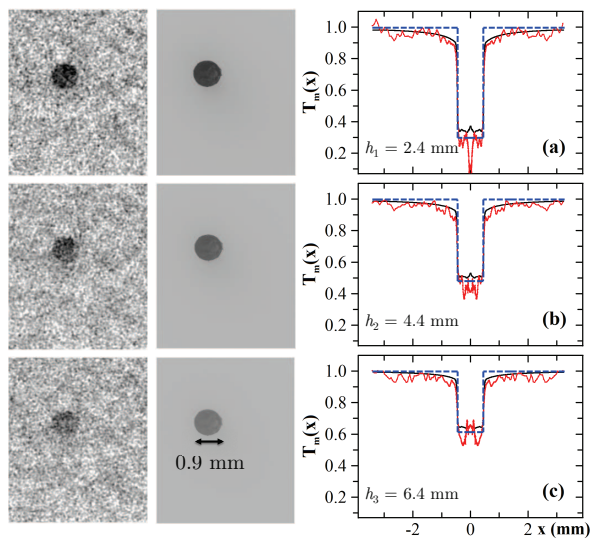} }
\caption{\label{Fig6} (Color online) Images of a small absorbing sphere (diam = $900\, \mu$m) inside a turbid solution with $l^*=6.7$mm at $h_1=2.4$mm (a), $h_2=4.4$mm (b) and $h_3=6.4$mm (c), obtained with DGI (left column) and standard imaging (central column). The right column plots the radial profiles averaged over the azimuthal angle: DGI (red curve), standard imaging (black curve), theoretical model (blue dashed line).}
\end{figure}
In this letter we have shown that DGI can be profitably used in a backscattering configuration for the imaging of small absorbing objects immersed in a turbid medium, in proximity of its surface. Spurred by the recent debate about the potentiality of GI for the imaging of objects in the presence of turbulence or scattering \cite{Meyers-Shapiro}, we have quantitatively compared DGI with a standard imaging method. Our results show that the two techniques perform almost identically and are equally affected by the presence of multiple scattering when the object is deeply immersed in the medium ($h\geq l^*$). This feature demonstrates that GI is not \emph{immune} from multiple scattering, exactly as it happens for GI when there is turbulence between the beam splitter and the object or the CCD \cite{Meyers-Shapiro}. However, there are situations where backscattering DGI may turn out to be very convenient, such as for example in biomedical tissue imaging for the early detection of pigmented skin lesions. In these cases, the existing optical methods \cite{Imaging_turbid} are either too qualitative (such as epiluminescence imaging or dermoscopy \cite{Epi}) or rather complex and expensive as the ones based on Optical Coherence- \cite{Bouma_OCT} and Diffuse Optical-Tomography \cite{DOT_Yodh} or Diffuse Reflection-correlation spectroscopy \cite{Yodh}. We therefore believe that backscattering DGI has the potentialities to become, in the next future, a valid imaging tool alternative or complementary to the current state of the art imaging techniques.

\end{document}